# Coupling of COAMPS™ and WAVEWATCH with Improved Wave Physics

Pat Fitzpatrick[1], Gueorgui Mostovoi[1], Yongzuo Li[1], Matt Bettencourt[2] and Shahrdad Sajjadi[2]

Mississippi State University[1]
Engineering Research Center
Bldg 1103, Room 233
Stennis Space Center, Mississippi 39529
Tel: 228-688-1157
Email: fitz@erc.msstate.edu

**Abstract**

The Model Coupling Executable Library (MCEL), developed at the University of Southern Mississippi's Center of Higher Learning, has been successfully used to couple the Coupled Ocean/Atmospheric Mesoscale Prediction System (COAMPS™) and the ocean wave model WAVEWATCH. An example of its application is shown for Hurricane Gordon, showing that two-way coupling results affects boundary layer physics differently than one-way coupling --- in this case, resulting in larger $z_o$ and, consequently, larger surface fluxes and a more intense hurricane. However, since analyzing MCEL is difficult because the wave physics is inaccurate, improvements to the wave algorithms are also part of the deliverables. A new analytical expression for the wind/wave growth factor has been derived based on normal modes analysis and rapid distortion theory valid for all wave regimes except for tropical cyclone conditions. This new algorithm is validated against a numerical simulation of the Reynolds-stress transport equations and matches well. In contrast, other wave growth expression used in ocean models like the WAve Model (WAM) and WAVEWATCH do not produce the same results, with larger wave growth values peaking at smaller wave age values. These differences are attributed to the application of curve fitting by the other algorithms, while the new formulation is an analytical expression derived from first principles and includes factors missing in previous schemes such as turbulent interaction. If the Reynolds-stress transport equations solutions are reasonably accurate, it indicates that all the previous wave growth schemes, including WAVEWATCH, have serious deficiencies. Another unique result from this work includes a second analytical wave growth formulation valid for tropical cyclone conditions.

---

[2] University of Southern Mississippi

An unexpected problem occurred with WAVEWATCH when it was discovered the roughness values are often one to two orders of magnitude too large. To circumvent this problem, the algorithm of Nordeng has been coded to compute roughness length for WAVEWATCH. This algorithm is a complicated iterative procedure involving integral expressions where turbulent stress, wave-induced stress, roughness length, and wave growth must converge.

## 1. Introduction

A major component of next-generation operational models includes coupling interaction between meteorology and wave models. However, software tools to facilitate coupling are currently unavailable, requiring major rewrites of both models to include interactions during time steps. Wave models also suffer from inaccurate forcing terms, often empirically derived and not based on physics. This work performs the following tasks: 1) Couples the atmospheric model COAMPS™ with the National Oceanic and Atmospheric Administration wave model WAVEWATCH using MCEL; 2) Shows an example of this coupling tool with a simulation of Hurricane Gordon, including how the intensity varies between one-way and two-way coupling with MCEL; 3) Develops new wave growth algorithms based on physics, not data fitting, for all wave age classes, validated against a boundary layer turbulence model; 4) Develops an alternative formulation for wave age roughness length $z_o$, as our research found serious errors in WAVEWATCH's $z_o$ formulation; and 5) Compares the new wave growth and $z_o$ schemes to other wave growth formulations used in WAM and WAVEWATCH.

## 2. MCEL and an Example Application

MCEL, developed at the University of Southern Mississippi's Center of Higher Learning, uses a data flow approach to model coupling where the communication is handled via the Common Object Request Broker Architecture (CORBA). In this approach, a central server is responsible for storing and passing information. The numerical models, or clients, are responsible for storing data into the server. Once a request is made for a set of data, the data flows from the server through a series of filters and to the client. These filters modify the data into a form that can be used by the clients, such as performing interpolation between the two different model grids or computing physical terms. In this manner, little modification of the model source code is required other than including filter subroutines, and both models can be synchronized for their respective time steps.

Coupling between the atmospheric model and wave model is achieved through two-way exchanges of 10-m wind speed (from COAMPS to WAVEWATCH) and $z_o$ (from WAVEWATCH to COAMPS). This interaction is important because $z_o$ affects the atmospheric boundary layer's heat fluxes, moisture fluxes, and wind stress, while wind forces wave growth (which is itself a function of wave age and $z_o$). These relationships have complicated nonlinear feedbacks, and require iterative procedures that will be discussed later. To show the usefulness of the MCEL interface for atmosphere-wave model coupling, a tropical cyclone simulation for Hurricane Gordon (2000) is performed for one-way and two-way coupling. Figures 1 and 2 show $z_o$ and latent heat values, respectively, for both simulations. The two-way coupling results in larger $z_o$ values, and as a consequence, larger latent heat flux values. The simulated hurricane responds with a central pressure 3-5 mb deeper, and wind speeds 5-10 knots stronger (not shown), which more closely matched observations. Gordon was a weak hurricane (Category 1, and sheared), and a "classic" hurricane will result in more dramatic results. MCEL's potential is obvious, as not only did it allow easy model coupling, but also facilitated straightforward diagnosis of coupling issues.

3.    **Wave Growth Physics**

Since Miles [1] first published the mechanism and a theoretical expression for surface wave generation by wind, much work has been done to formulate a proper wind-wave interaction source term into ocean wave models. Many of these algorithms involve parameterizing the non-dimensional wave growth term $\beta$. Formulations are derived by four mechanisms: analytical expressions such as done by Miles [1] and Sajjadi [2]; empirical fits to the Miles' Rayleigh equation such as performed by Janssen [3] and used in ocean models like WAM; empirical fits to limited observation studies which are of questionable accuracy and usefulness (see [4] for a review) but still used in wave models like WAM and WAVEWATCH; and empirical fits to numerical model results of turbulent boundary layer flow over a moving gravity surface waves as done by Burgers and Makin [5] and the default scheme used in WAVEWATCH. As will be seen, these equations give surprisingly different answers. Much of this uncertainty stems from the use of curve-fitting, which avoids physical understanding of wave growth, and other pitfalls of empiricism (overfitting, applying these equations outside their sampling regime, validity of the least squares approach, applicability to all wave ages, etc.). Furthermore, while the Miles work is commendable for its physics-based algorithms, a major inadequacy is the neglect of any interaction between the waves and small-scale air turbulence, which is known to increase wave growth.

Therefore, since analyzing MCEL is difficult if the wave physics is inaccurate, a secondary approach in this work is to derive an analytical physics-based $\beta$ expression, valid for all wave ages, and includes turbulent and wind shear interaction. Based on normal modes analysis and rapid distortion theory, this work has resulted in the submission of two papers for peer-review ([4] and [6]), which yield the following results:

$$\beta_{saj} = \beta_{turb} + \beta_{crit}$$
$$\beta_{turb} = 5\kappa^2 l_\circ, \quad \beta_{crit} = 2.5\pi \hat{W} l_\circ^4 \left[ 1 - \frac{1}{l_\circ^2}\left(4 - \frac{\pi^2}{3}\right) \right]$$
$$l_\circ = -0.5772 - \ln \hat{W}$$
$$\hat{W} = \kappa z_\circ \left[\frac{2u_*}{c}\right]^2 \exp\left[\frac{c}{2u_*}\right]$$
(1)

where $\kappa$ is the Von Karman constant 0.4, $c$ is the peak phase speed, and $u_*$ is the friction velocity. $\beta_{turb}$ corresponds to the effect of turbulence on wave growth, and $\beta_{crit}$ is closely related to the critical layer mechanism discovered by Miles. Equation 1 is valid for wave ages $\frac{c}{u_*} > 5$.

For very windy conditions when $\frac{c}{u_*} < 5$, such as in a hurricane, the following expression has been derived:

$$\beta_{hurr} = \left[1 + \frac{\rho_{air}}{\rho_{water}} \frac{kF}{c^3(ak)^2}\right]^{-1} F, \qquad (2)$$

where $\rho$ is density, $k$ is wavenumber, $a$ is a wave steepness parameter, and $F$ is a complicated turbulent flux parameter. Turbulent numerical models show that $\beta$ saturates when $\frac{c}{u_*} < 5$, but to the best of our knowledge, this is the first analytical expression for this regime.

Equation 1 is validated for constant $z_\circ$ against a numerical simulation of the Reynolds-stress transport equations. These results are also compared against the Janssen formulation $\beta_{jan}$, which is an empirical fit to Miles' Rayleigh equation, as well as Miles approximate solution $\beta_{miles}$. As shown in Figure 3, $\beta_{saj}$ matches the numerical simulation well, while the other expressions are too large, and peak at smaller values of $\frac{c}{u_*}$. These differences can be attributed to the use of Rapid Distortion Theory, accounting for turbulence in $\beta_{saj}$, and other reasons discussed in [4].

**4.    Comparisons of $\beta$ for Wave-Age Based $z_\circ$ and Problems with WAVEWATCH $z_\circ$**

For true comparisons, $z_\circ$ should not be held constant but allowed to vary with wind speed and wave age. An unexpected problem occurred with WAVEWATCH when it was discovered the $z_\circ$ values are often *one to two orders of magnitude too large*, thereby stalling the model coupling deliverable of this project. . After investigating, it was discovered that curve fitting had been applied to a turbulent simulation consisting of drag coefficient values with large scatter [7], but this problem requires more in-depth analysis. Perhaps this problem has never been noticed because the $z_\circ$ values have not been analyzed or used for coupling before. To circumvent this problem, the algorithm of Nordeng [8] is used to compute $z_\circ$. This algorithm is a complicated iterative procedure involving integral expressions where turbulent stress $\tau_{turb}$, wave-induced stress $\tau_{wave}$, $u_*$, and $\beta$ must converge. However, it gave the most realistic $z_\circ$ values. Figure 4 shows plots of the WAVEWATCH $\beta$ using the Nordeng scheme, as well as $\beta_{saj}$, $\beta_{jan}$, and $\beta_{miles}$ for a wind speed of 10 ms$^{-1}$ and peak phase speeds ranging from 2 to 12 ms$^{-1}$. As can be seen, the results are very different, indicating that more research is needed on wave-wind interaction. However, if the Reynolds-stress transport equations solutions are reasonably accurate, it indicates that all previous $\beta$ schemes, including WAVEWATCH, have serious deficiencies.

5.     **Summary**

MCEL has been successfully used to couple COAMPS and WAVEWATCH. An example of its application is shown for Hurricane Gordon, showing that two-way coupling results affects boundary layer physics differently than one-way coupling --- in this case, resulting in larger $z_o$ and, consequently, larger surface fluxes and a more intense hurricane. A new analytical expression for the wind/wave growth factor has been derived based on normal modes analysis and rapid distortion theory valid for all wave regimes except for tropical cyclone conditions. This new algorithm is validated for constant roughness length against a numerical simulation of the Reynolds-stress transport equations and matches well. In contrast, other wave growth expression used in ocean models like WAM and WAVEWATCH do not produce the same results, with larger wave growth values peaking at smaller wave age values. These differences are attributed to the application of curve fitting by the other algorithms, while the new formulation is an analytical expression derived from first principles, and includes factors missing in previous schemes such as turbulent interaction. If the Reynolds-stress transport equations solutions are reasonably accurate, it indicates that all the previous wave growth schemes, including WAVEWATCH, have serious deficiencies. Another unique result from this work includes a wave growth formulation valid for tropical cyclone conditions.

An unexpected problem occurred with WAVEWATCH when it was discovered the roughness lengths values are often one to two orders of magnitude incorrect. To circumvent this problem, the algorithm of Nordeng [8] has been coded to compute $z_o$ for WAVEWATCH. This algorithm is a complicated iterative procedure involving integral expressions where turbulent stress, wave-induced stress, roughness length, and wave growth must converge.

**Acknowledgment**

This publication made possible through support provided by DoD High Performance Computing Modernization Program (HPCMP) Programming Environment and Training (PET) activities through Mississippi State University under the terms of Contract No. N62306-01-D-7110. Views, opinions, and/or findings contained in this report are those of the authors and should not be construed as an official DoD position, policy or decision unless so designated by other official documentation and no official endorsement should be inferred.

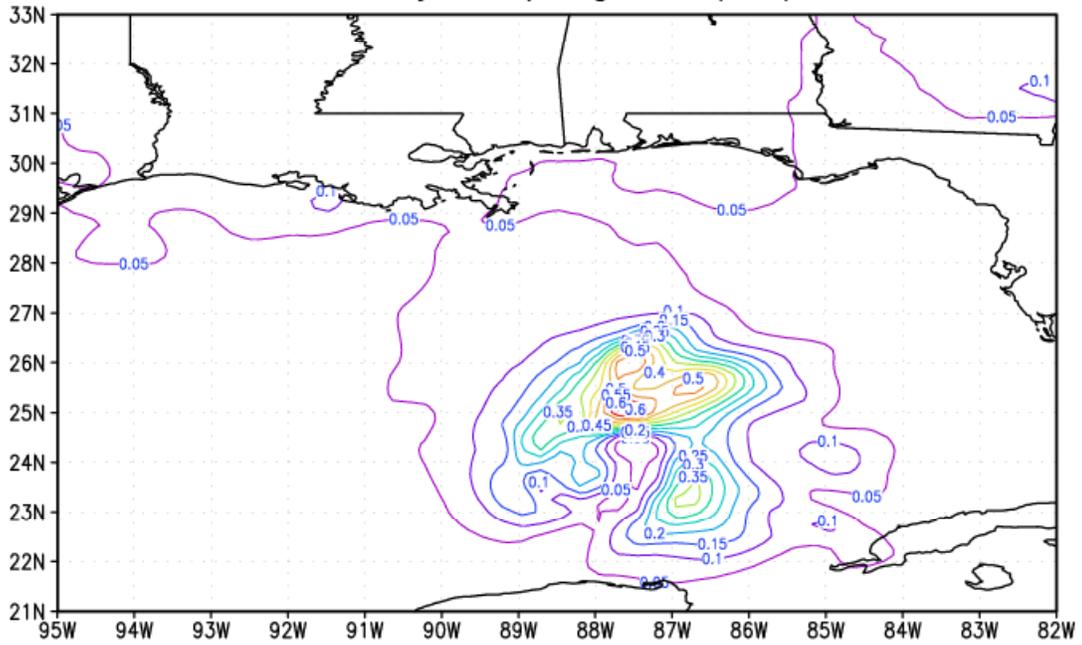
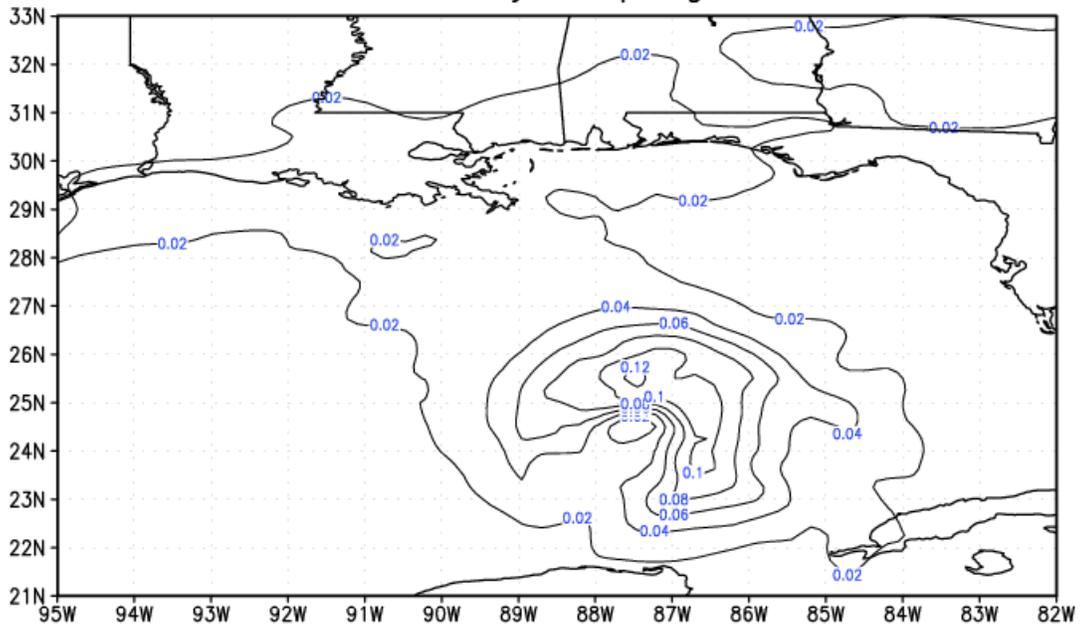

Figure 1. Plots of $z_0$ (cm) are shown for two-way coupling (top) and one-way coupling (bottom) for Hurricane Gordon (2000) using MCEL software. The two-way coupling produces larger $z_0$ values 18 h into the simulation.

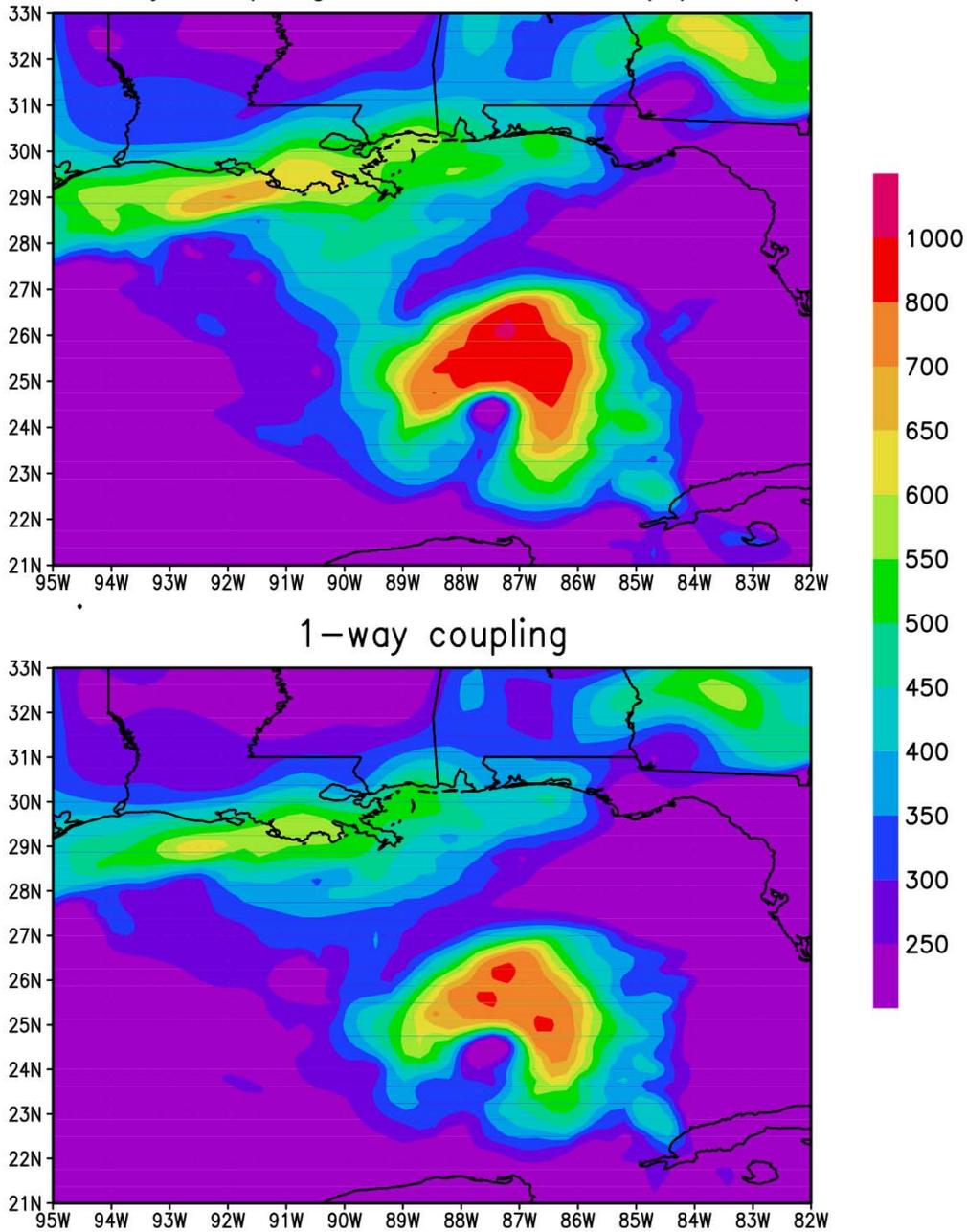

Figure 2. As in Figure 1, but for latent heat flux. The larger $z_\circ$ values increase the surface fluxes, resulting in a stronger hurricane.

**Wave growth $\beta$ (y-axis) vs. wave-age $c/u_*$ (x-axis) for constant $kz_\circ = 10^{-4}$**

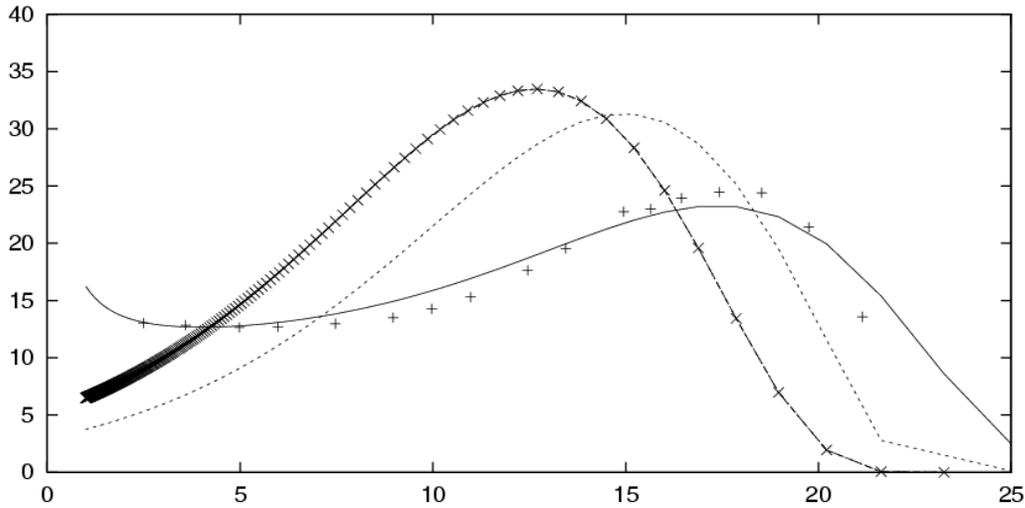

Figure 3. Wave growth parameter $\beta$ versus wave age $c/u_*$ for constant $z_\circ$. xxxx., Janssen's formulation [3]; dashed line, Miles' analytical solution [1]; solid line, new analytical solution from Equation 1 [4]; ++++, numerical solutions of Reynold-stress transport equations. Equation 1 matches the numerical simulation well, while the others are too large, and peak at smaller $c/u_*$.

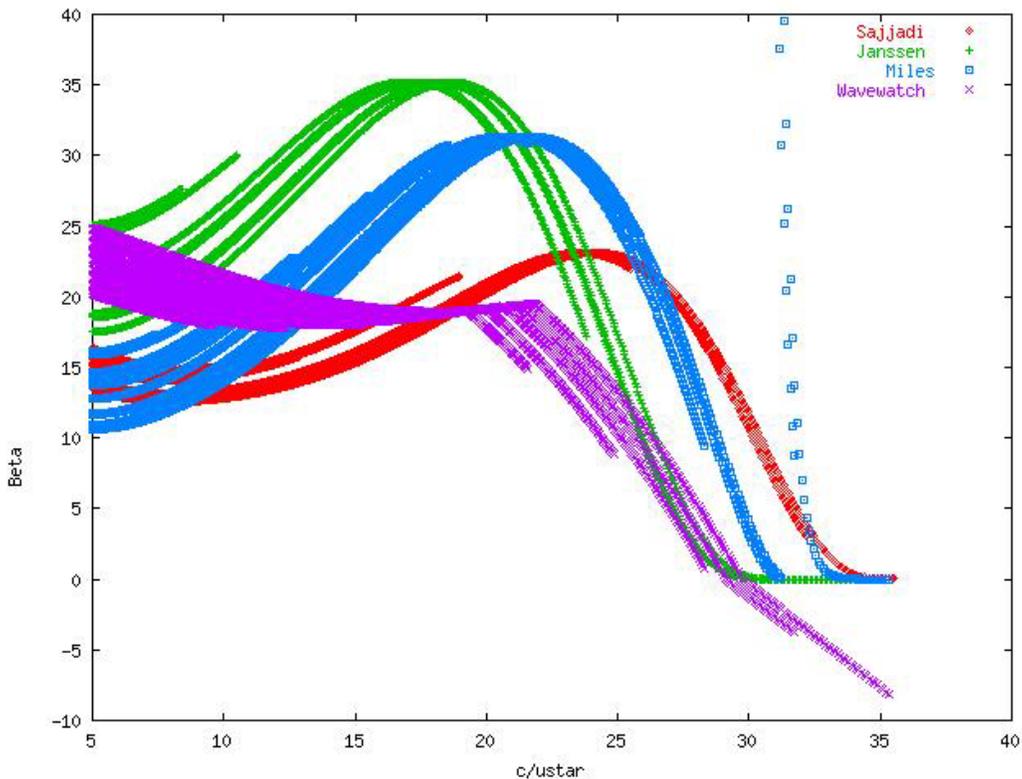

Figure 4. Solutions of $\beta$ from Miles (blue, [1]), Janssen (green, [3]), WAVEWATCH (purple, [5]) and Equation 1 (red, [4]) for a 10 ms$^{-1}$ wind and peak phase speeds ranging from 2-12 ms$^{-1}$ using the wave-age $z_\circ$ algorithm of Nordeng [8]. The different solutions certainly indicate more research is needed on this subject.